\documentclass[pre,aps,12pt]{revtex4}
\usepackage{graphicx}

\begin{document}

\title{Two-dimensional spin models with macroscopic degeneracy}
\author{D.~V.~Dmitriev}
\email{dmitriev@deom.chph.ras.ru}
\author{V.~Ya.~Krivnov}
\date{}

\begin{abstract}
We consider a class of anisotropic spin-$\frac{1}{2}$ models with
competing ferro- and antiferromagnetic interactions on
two-dimensional Tasaki and kagome lattices consisting of corner
sharing triangles. For certain values of the interactions the
ground state is macroscopically degenerated in zero magnetic
field. In this case the ground state manifold consists of isolated
magnons as well as the bound magnon complexes. The ground state
degeneracy is estimated using a special form of exact wave
function which admits arrow configuration representation on
two-dimensional lattice. The comparison of this estimate with the
result for some special exactly solved models shows that the used
approach determines the number of the ground states with
exponential accuracy. It is shown that the main contribution to
the ground state degeneracy and the residual entropy is given by
the bound magnon complexes.
\end{abstract}

\maketitle

\affiliation{Institute of Biochemical Physics, RAS, Kosygin str.
4, 119334, Moscow, Russia.}

\section{Introduction}

Quantum magnets on geometrically frustrated lattices have
attracted much interest in last year
\cite{Diep,Lacroice,ModernPhysics}. An important class of these
systems includes lattices with magnetic ions located on vertices
of connected triangles. For special relation between different
exchange interactions such systems have a dispersionless
one-magnon band \cite{ModernPhysics}. There is a wide class of the
frustrated quantum antiferromagnets (AF) in which a flat band
exists \cite{Derg,Zhitomirsky,Tsun,Derg1,Zhit,Schulen}. It
includes the kagome and pyrochlor antiferromagnets. An interesting
one-dimensional example of the flat-band antiferromagnet is the
saw-tooth chain. The one-magnon flat-band leads to an existence of
exact localized many-magnon states. In the antiferromagnetic
flat-band models these states form the ground state manifold in
the saturation magnetic field and the degeneracy grows
exponentially in the thermodynamic limit giving a finite residual
entropy \cite{Derg,Zhitomirsky}.

Another class of the frustrated quantum models with exact
localized multi-magnon states are the models with competing ferro-
and antiferromagnetic interactions (F-AF models). The
zero-temperature phase diagram of these models has different
phases depending on the ratio of the ferromagnetic and
antiferromagnetic interactions. At the critical value of this
ratio corresponding to a phase boundary the model has massively
degenerated ground state manifold. One example of the F-AF systems
is the same saw-tooth chain at the critical value of the
frustration parameter $\alpha =\frac{J_{2}}{\left\vert
J_{1}\right\vert }=\frac{1}{2}$, where $J_{2} $ and $J_{1}$ are AF
and F interactions of the basal-basal and basal-apical spins,
correspondingly \cite{Dmitriev}. It is worth noting that this
model describes the magnetic molecule $Fe_{10}Gd_{10}$ frustration
parameter of which is close to the critical value \cite{Gd,DKRS}.
The ground state spin of this molecule $S=60$ is one of the
largest spin of a molecule.

The main difference between AF and F-AF models is the additional
bound magnon complexes which are exact ground states at the
critical value of the frustration parameter together with the
independent localized magnons. It leads to the macroscopical
degeneracy of the ground state in zero-magnetic field and the
residual entropy is larger than that for the AF models. The
residual entropy in zero-magnetic field leads to an efficient
magnetic cooling \cite{cooling,Dmitriev} which is important from
the practical point of view.

The saw-tooth model and special type of kagome-like stripes
\cite{kagomestripe} in the critical point are solved exactly in
contrast with two-dimensional F-AF models. At the same time the
degeneracy of the AF models and the residual entropy in the
saturation field can be found by mapping the system of isolated
localized magnons onto the classical lattice gas of hard-core
particles \cite{Derg}. However, such mapping does not work for the
F-AF models due to the existence of the bound magnon complexes.

In recent papers \cite{3-colors,3-colors1} the $s=\frac{1}{2}$ XXZ
model on the two-dimensional lattices consisting of triangles has
been studied for a special value of the ratio of the Ising $J_{z}$
to the antiferromagnetic transverse interaction $J$
($\frac{J_{z}}{J}=-\frac{1}{2}$) when the ground state has
macroscopic degeneracy in zero-magnetic field. It was shown
earlier in Refs.\cite{Dmitriev1,Johann} that this model has both
localized many-magnon states and bound magnon complexes and all of
them belong to the ground state manifold. It was shown in
\cite{3-colors,3-colors1} that the ground state can be described
by `quantum three-coloring wave functions'. This approach is based
on a representation of the exact ground states in terms of the
wave functions denoted as colors on sites of the lattice and each
site can be painted in one of three colors, while the nearest
sites can not be the same color. We note, however, that the `three
colors representation' is applicable to the models with equal
interactions between all nearest spins only.

In the present paper we extent this approach to a class of the
spatially anisotropic F-AF two-dimensional models with the
macroscopically degenerated ground state. Our main task is the
calculation of the total number of the ground states, $W$. The
macroscopic degeneracy means that $W$ grows exponentially in the
thermodynamic limit, i.e $W\sim \xi ^{N}$ ($N$ is number of spins)
and we focus on the evaluation of $\xi $ and the corresponding
entropy per spin $\mathcal{S=}\ln \xi $. In other words, we will
estimate the exponential factor of the total degeneracy.

The paper is organized as follows. In Section II we establish
necessary conditions on the Hamiltonian parameters for which the
F-AF models have a massive degenerate ground state and consider
some special cases of these parameters. In Section III we
represent the exact ground state wave function of the system as a
product of exact wave functions of the connected triangles and we
give the lattice gas representation of this function in terms of
configurations of arrows on the bond between two spins. As an
example we apply this approach to the F-AF saw-tooth chain in the
critical points and compare obtained results with the exact ones.
In Section VI we estimate the total number of the degenerate
ground states for the two-dimensional F-AF Tasaki and kagome
lattices and compare the obtained results with exact
diagonalization of finite systems. The conclusions are summarized
in Section V.

\section{Anisotropic F-AF models with macroscopical ground state degeneracy}

In this Section we find the general form of the Hamiltonian of the
$s=\frac{1}{2}$ F-AF model consisting of corner-sharing triangles
for which the model becomes flat-band one with highly degenerate
ground state. The Hamiltonian of such model can be written as a
sum of local Hamiltonians acting on triangles:
\begin{equation}
H=\sum H_{i}  \label{H}
\end{equation}

\begin{figure}[tbp]
\includegraphics[width=3in,angle=0]{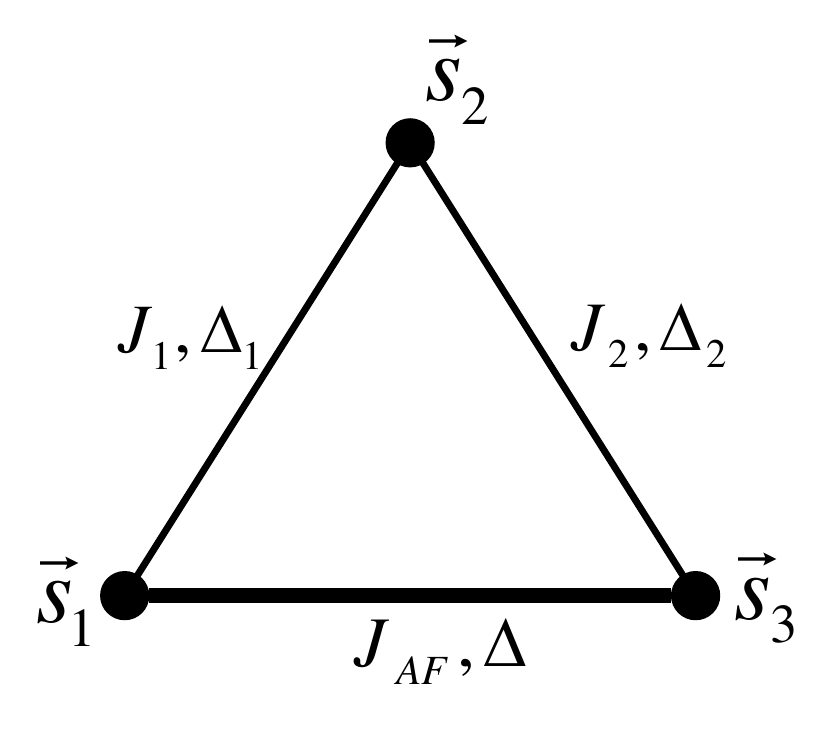}
\caption{Local Hamiltonian on triangle.}
\label{Fig_triangle}
\end{figure}

The most general form of such local Hamiltonian commuting with the
total $S_{tot}^{z}$ is
\begin{eqnarray}
H_{1} &=&-J_{1}\left( s_{1}^{x}s_{2}^{x}+s_{1}^{y}s_{2}^{y}+\Delta
_{1}(s_{1}^{z}s_{2}^{z}-\frac{1}{4})\right) -J_{2}\left(
s_{2}^{x}s_{3}^{x}+s_{2}^{y}s_{3}^{y}+\Delta _{2}(s_{2}^{z}s_{3}^{z}-\frac{1%
}{4})\right)  \nonumber \\
&&+J_{AF}\left( s_{1}^{x}s_{3}^{x}+s_{1}^{y}s_{3}^{y}+\Delta
(s_{1}^{z}s_{3}^{z}-\frac{1}{4})\right)  \label{H1}
\end{eqnarray}%
where the first two terms describe two ferromagnetic interactions
and the third one describes antiferromagnetic interaction of spins
on the triangle (see Fig.\ref{Fig_triangle}). The constants in
(\ref{H1}) are chosen so that the energy of the ferromagnetic
state on triangle with $S^{z}=\pm \frac{3}{2}$ is zero. Further we
scale the energy by the normalization $J_{AF}=1$.

When $J_{1}$ and $J_{2}$ are large the ground state of $H$ is
ferromagnetic ($S_{tot}^{z}$ $=\frac{N}{2}$) with zero energy. At
the transition points on the phase boundaries between the
ferromagnetic and another phase (antiferromagnetic or
ferrimagnetic \cite{ferri}) the ground state energy remains zero
as well as the energy of $H_{1}$ with $S^{z}=\pm \frac{3}{2}$. As
was shown in Ref.\cite{Dmitriev} for the saw-tooth chain with
equal ferromagnetic bonds ($J_{1}=J_{2}$, $\Delta _{1}=\Delta
_{2}$) at the transition points the ground state is
macroscopically degenerated and the spectrum of each local
Hamiltonian (\ref{H1}) consists of two excited states with
$S^{z}=\pm \frac{1}{2}$ and six ground states (two states with
$S^{z}=\pm \frac{3}{2}$ and four states with $S^{z}=\pm
\frac{1}{2}$). The same properties of local Hamiltonian remains
for the saw-tooth chain with isotropic interactions ($\Delta
_{i}=1$) but different ferromagnetic bonds $J_{1}\neq J_{2}$
\cite{kosoi}. The local Hamiltonian which has the above spectrum
can be written as a sum of two projectors on two excited states
with arbitrary numerical factors. Therefore, we need to write down
the general form of these excited states.

The basis for the system of three spins-$\frac{1}{2}$ in the
sector with $S^{z}=\frac{1}{2}$ contains three states: $\left\vert
\downarrow \uparrow \uparrow \right\rangle ,\left\vert \uparrow
\downarrow \uparrow \right\rangle ,\left\vert \uparrow \uparrow
\downarrow \right\rangle $ (and similar for the sector with
$S^{z}=-\frac{1}{2}$). Therefore, any state with
$S^{z}=\frac{1}{2}$ can be parameterized by two independent
parameters. It is convenient to choose these parameters as two
angles $\alpha $ and $\beta $, so that the excited states with
$S^{z}=\pm \frac{1}{2}$ are written in the
following form:%
\begin{eqnarray}
\chi _{-1/2} &=&\sin \beta \left\vert \uparrow \downarrow \downarrow
\right\rangle -\sin (\alpha +\beta )\left\vert \downarrow \uparrow
\downarrow \right\rangle +\sin \alpha \left\vert \downarrow \downarrow
\uparrow \right\rangle  \label{excited0} \\
\chi _{1/2} &=&\sin \beta \left\vert \downarrow \uparrow \uparrow
\right\rangle -\sin (\alpha +\beta )\left\vert \uparrow \downarrow
\uparrow \right\rangle +\sin \alpha \left\vert \uparrow \uparrow
\downarrow \right\rangle   \label{excited}
\end{eqnarray}

Then, the ground state manifold contains two ferromagnetic states
$\left\vert \downarrow \downarrow \downarrow \right\rangle $,
$\left\vert \uparrow \uparrow \uparrow \right\rangle $ and four
states with $S^{z}=\pm \frac{1}{2}$ orthogonal to the states
$\chi_{\pm 1/2}$, which can be chosen as:
\begin{eqnarray}
\phi _{-1/2} &=&\cos \alpha \left\vert \uparrow \downarrow \downarrow
\right\rangle +\left\vert \downarrow \uparrow \downarrow \right\rangle +\cos
\beta \left\vert \downarrow \downarrow \uparrow \right\rangle  \label{states0} \\
\phi _{1/2} &=&\cos \alpha \left\vert \downarrow \uparrow \uparrow
\right\rangle +\left\vert \uparrow \downarrow \uparrow
\right\rangle +\cos
\beta \left\vert \uparrow \uparrow \downarrow \right\rangle  \\
\varphi _{-1/2} &=&\sin \alpha \left\vert \uparrow \downarrow \downarrow
\right\rangle -\sin \beta \left\vert \downarrow \downarrow \uparrow
\right\rangle  \\
\varphi _{1/2} &=&\sin \alpha \left\vert \downarrow \uparrow \uparrow
\right\rangle -\sin \beta \left\vert \uparrow \uparrow \downarrow
\right\rangle   \label{states}
\end{eqnarray}

The interaction parameters of the local Hamiltonian (\ref{H1})
having the above spectrum is:
\begin{eqnarray}
\Delta  &=&\cos (\alpha +\beta ) \\
J_{1} &=&\frac{\sin \left( \alpha +\beta \right) }{\sin \alpha } \\
J_{2} &=&\frac{\sin \left( \alpha +\beta \right) }{\sin \beta } \\
J_{1}\Delta _{1} &=&\cos (\alpha +\beta )+\frac{\sin \beta }{\sin \alpha } \\
J_{2}\Delta _{2} &=&\cos (\alpha +\beta )+\frac{\sin \alpha }{\sin \beta }
\end{eqnarray}

Energy of the excited states $\chi _{\pm 1/2}$ on triangle is%
\begin{equation}
E=\cos (\alpha +\beta )+\frac{\sin \beta }{\sin \alpha }+\frac{\sin \alpha }{%
\sin \beta }
\end{equation}

Both angles $\alpha $ and $\beta $ can be analytically continued.
Since the parameters $J_{i}$ and $\Delta _{i}$ of the Hamiltonian
(\ref{H1}) are real, $\alpha $ and $\beta $ can be either both
real ($\Delta \leq 1$) or both imaginary ($\Delta >1$). For the
case of imaginary $\alpha $ and $\beta $ we can use the same
expressions for wave functions and Hamiltonian parameters just
substituting $\alpha \to i\alpha $ and $\beta \to i\beta $ (so
that, $\cos \to \cosh $ and $\sin \to \sinh $ everywhere):
\begin{eqnarray}
\Delta &=&\cosh (\alpha +\beta ) \label{Im1} \\
J_{1} &=&\frac{\sinh \left( \alpha +\beta \right) }{\sinh \alpha } \\
J_{2} &=&\frac{\sinh \left( \alpha +\beta \right) }{\sinh \beta } \\
J_{1}\Delta _{1} &=&\cosh (\alpha +\beta )+\frac{\sinh \beta }{\sinh \alpha }
\\
J_{2}\Delta _{2} &=&\cosh (\alpha +\beta )+\frac{\sinh \alpha
}{\sinh \beta } \label{Im2}
\end{eqnarray}

\subsection{Special cases}

Let us consider some special cases of the Hamiltonian $H$.

\textbf{The case }$\alpha =\beta $ corresponds to the Hamiltonian
(\ref{H1}) with $\Delta =\cos (2\alpha )$ and equal ferromagnetic
bonds ($\Delta _{1}=\Delta _{2}=\cos \alpha $\ and
$J_{1}=J_{2}=2\cos \alpha $). This case was studied in detail for
saw-tooth chain in Ref.\cite{Dmitriev1}.

\textbf{The case }$\alpha \to 0$, $\beta \to 0$. This case describes the
isotropic model ($\Delta =\Delta _{1}=\Delta _{2}=1$) with
\begin{eqnarray}
J_{1} &=&1+\frac{\beta }{\alpha } \\
J_{2} &=&1+\frac{\alpha }{\beta }
\end{eqnarray}%
so that $J_{1}$ and $J_{2}$ obey the relation:
$J_{1}^{-1}+J_{2}^{-1}=1$. This case was studied for saw-tooth
chain in detail in Ref.\cite{kosoi}.

\textbf{Special case }$\alpha +\beta =\frac{\pi }{2}$. In this
case $\Delta =0$ and the local Hamiltonian takes the form:
\begin{equation}
H_{1}=-\frac{1}{\sin \alpha }\left(
s_{1}^{x}s_{2}^{x}+s_{1}^{y}s_{2}^{y}+\cos (\alpha
)s_{1}^{z}s_{2}^{z}\right) -\frac{1}{\cos \alpha }\left(
s_{2}^{x}s_{3}^{x}+s_{2}^{y}s_{3}^{y}+\sin (\alpha
)s_{2}^{z}s_{3}^{z}\right) +\left(
s_{1}^{x}s_{3}^{x}+s_{1}^{y}s_{3}^{y}\right)
\end{equation}

Here for $\alpha =\beta =\frac{\pi }{4}$ we reproduce the special
case $\Delta _{1}=\Delta _{2}=1/\sqrt{2}$ considered in our paper
\cite{Dmitriev1} for the F-AF saw-tooth chain. This case is
specific and it will be discussed below in detail. In this case
the Hamiltonian (\ref{H1}) is
\begin{equation}
H_{S,1}=\left( s_{1}^{x}s_{3}^{x}+s_{1}^{y}s_{3}^{y}\right) -\sqrt{2}%
(s_{1}^{x}s_{2}^{x}+s_{1}^{y}s_{2}^{y})-s_{1}^{z}s_{2}^{z}-\sqrt{2}%
(s_{2}^{x}s_{3}^{x}+s_{2}^{y}s_{3}^{y})-s_{2}^{z}s_{3}^{z}  \label{sqrt2}
\end{equation}

In case $\alpha \to 0$ and $\beta \to \frac{\pi }{2}$, the model
has the isotropic ferromagnetic bond ($1-2$) slightly connected
with the third spin by the $XY$ type interactions:
\begin{equation}
H_{1}=-\mathbf{\vec{s}}_{1}\cdot \mathbf{\vec{s}}_{2}-\alpha \left(
s_{2}^{x}s_{3}^{x}+s_{2}^{y}s_{3}^{y}\right) +\alpha \left(
s_{1}^{x}s_{3}^{x}+s_{1}^{y}s_{3}^{y}\right)  \label{F12}
\end{equation}

\textbf{The case }$\alpha =\beta =\frac{\pi }{3}$. In this case
$J_{1}=$ $J_{2}=-1$, $\Delta _{1}=\Delta _{2}=\frac{1}{2}$,
$\Delta =-\frac{1}{2}$. After rotation in the $XY$ plane
$s_{n}^{x,y}\to (-1)^{n}s_{n}^{x,y}$ the local Hamiltonian takes
the symmetric form
\begin{equation}
H_{1}=\left( s_{1}^{x}s_{2}^{x}+s_{1}^{y}s_{2}^{y}-\frac{1}{2}%
s_{1}^{z}s_{2}^{z}\right) +\left( s_{2}^{x}s_{3}^{x}+s_{2}^{y}s_{3}^{y}-%
\frac{1}{2}s_{2}^{z}s_{3}^{z}\right) +\left(
s_{1}^{x}s_{3}^{x}+s_{1}^{y}s_{3}^{y}-\frac{1}{2}s_{1}^{z}s_{3}^{z}\right)
\label{pi/3}
\end{equation}

Exact ground state degeneracy for this model on saw-tooth chain
was calculated in Ref.\cite{Dmitriev1}. Later this model was
studied on kagome lattice in Ref.\cite{3-colors,3-colors1} using
three-coloring representation of the ground state manifold.

\textbf{The `imaginary' case} (\ref{Im1}-\ref{Im2}) with $\alpha
\gg 1$ and $\beta \gg 1$. The Hamiltonian in this case becomes
with the exponential accuracy (corrections of the order $\sim
e^{-\alpha }$or $e^{-\beta }$):
\begin{equation}
H_{1}=\left( s_{1}^{z}s_{3}^{z}-s_{1}^{z}s_{2}^{z}-s_{2}^{z}s_{3}^{z}\right)
-2e^{-\alpha }\left( s_{1}^{x}s_{2}^{x}+s_{1}^{y}s_{2}^{y}\right)
-2e^{-\beta }\left( s_{2}^{x}s_{3}^{x}+s_{2}^{y}s_{3}^{y}\right)
\end{equation}

The Ising part of the Hamiltonian $\left(
s_{1}^{z}s_{3}^{z}-s_{1}^{z}s_{2}^{z}-s_{2}^{z}s_{3}^{z}\right) $
has the ground state degeneracy ($3^{n}+1$) on saw-tooth chain
\cite{Dmitriev1}. The two last terms are $XY$ perturbations which
partially split the degeneracy. The split excited states form
multi-scale energy hierarchy \cite{Dmitriev1}.

\textbf{The `imaginary' case} with $\alpha \ll 1$. In this case
the model has the isotropic ferromagnetic bond ($1-2$) slightly
connected with the third spin by $XXZ$ type with opposite signs of
interactions between spins ($2,3$) and ($1,3$):
\begin{equation}
H_{1}=-\mathbf{\vec{s}}_{1}\cdot \mathbf{\vec{s}}_{2}-\frac{\alpha }{\sinh
\beta }\left( s_{2}^{x}s_{3}^{x}+s_{2}^{y}s_{3}^{y}+\cosh (\beta
)s_{2}^{z}s_{3}^{z}\right) +\frac{\alpha }{\sinh \beta }\left(
s_{1}^{x}s_{3}^{x}+s_{1}^{y}s_{3}^{y}+\cosh (\beta )s_{1}^{z}s_{3}^{z}\right)
\end{equation}

In the limit $\beta \gg 1$ only the Ising term survives in the 2-3
and 1-3 interactions and the model reduces to:
\begin{equation}
H_{1}=-\mathbf{\vec{s}}_{1}\cdot \mathbf{\vec{s}}_{2}-\alpha
s_{2}^{z}s_{3}^{z}+\alpha s_{1}^{z}s_{3}^{z}
\end{equation}

The latter model is similar to the model (\ref{F12}) but with the
Ising type of small perturbations.

\section{Arrow configuration approach for the ground state with macroscopic
degeneracy}

The exact ground state degeneracy of the F-AF saw-tooth chain at
the critical points has been found in \cite{Dmitriev,Dmitriev1}.
However, the consideration in \cite{Dmitriev,Dmitriev1} was relied
essentially on the one-dimensionality of the model and the methods
used in \cite{Dmitriev,Dmitriev1} are not applicable for the
two-dimensional case. In this Section we will use another approach
for the estimate of the ground state degeneracy and as an example
we consider the F-AF saw-tooth chain.

First, we notice that four states $\phi _{\pm 1/2}$ and $\varphi
_{\pm 1/2}$ (\ref{states0}-\ref{states}) can be linearly
transformed to four functions $\psi _{\pm 1/2}=\phi _{\pm
1/2}-i\varphi _{\pm 1/2}$ and $\psi _{\pm 1/2}^{\ast }=\phi _{\pm
1/2}+i\varphi _{\pm 1/2}$:
\begin{eqnarray}
\psi _{-1/2} &=&\left( e^{-i\alpha }s_{1}^{+}+s_{2}^{+}+e^{i\beta
}s_{3}^{+}\right) \left\vert \downarrow \downarrow \downarrow \right\rangle
\label{function01} \\
\psi _{-1/2}^{\ast } &=&\left( e^{i\alpha
}s_{1}^{+}+s_{2}^{+}+e^{-i\beta }s_{3}^{+}\right) \left\vert
\downarrow \downarrow \downarrow \right\rangle
\\
\psi _{1/2} &=&\left( e^{-i\alpha
}s_{2}^{+}s_{3}^{+}+s_{1}^{+}s_{3}^{+}+e^{i\beta }s_{1}^{+}s_{2}^{+}\right)
\left\vert \downarrow \downarrow \downarrow \right\rangle  \\
\psi _{1/2}^{\ast } &=&\left( e^{i\alpha
}s_{2}^{+}s_{3}^{+}+s_{1}^{+}s_{3}^{+}+e^{-i\beta }s_{1}^{+}s_{2}^{+}\right)
\left\vert \downarrow \downarrow \downarrow \right\rangle   \label{function1}
\end{eqnarray}

We look for the wave function of the ground state in the form:%
\begin{equation}
\Psi =\prod (1+e^{i\varphi _{m}}s_{m}^{+})\left\vert \downarrow \downarrow
\ldots \downarrow \right\rangle  \label{function}
\end{equation}%
where the product in Eq.(\ref{function}) is taken over all sites.
This wave function contains all possible values of
$S_{tot}^{z}=-\frac{1}{2}N\ldots \frac{1}{2}N$ (it does not
preserve total $S_{tot}^{z}$). The wave function (\ref{function})
resembles three-coloring approach for the construction of the
ground state \cite{3-colors,3-colors1}.

Let us consider the the part of the wave function (\ref{function})
corresponding to one isolated triangle. Its wave function
\begin{equation}
\Psi_{123}=(1+e^{i\varphi _{1}}s_{1}^{+})(1+e^{i\varphi
_{2}}s_{2}^{+})(1+e^{i\varphi _{3}}s_{3}^{+})\left\vert \downarrow
\downarrow \downarrow \right\rangle
\end{equation}%
is a superposition of both ferromagnetic states $\left\vert
\downarrow \downarrow \downarrow \right\rangle $, $\left\vert
\uparrow \uparrow \uparrow \right\rangle $ and the following wave
functions with $S^{z}=\pm \frac{1}{2}$:
\begin{eqnarray}
\psi _{-1/2} &=&\left( e^{i\varphi _{1}}s_{1}^{+}+e^{i\varphi
_{2}}s_{2}^{+}+e^{i\varphi _{3}}s_{3}^{+}\right) \left\vert \downarrow
\downarrow \downarrow \right\rangle \\
\psi _{1/2} &=&e^{i(\varphi _{1}+\varphi _{2}+\varphi _{3})}\left(
e^{-i\varphi _{1}}s_{2}^{+}s_{3}^{+}+e^{-i\varphi
_{2}}s_{1}^{+}s_{3}^{+}+e^{-i\varphi _{3}}s_{1}^{+}s_{2}^{+}\right)
\left\vert \downarrow \downarrow \downarrow \right\rangle
\end{eqnarray}

\begin{figure}[tbp]
\includegraphics[width=5in,angle=0]{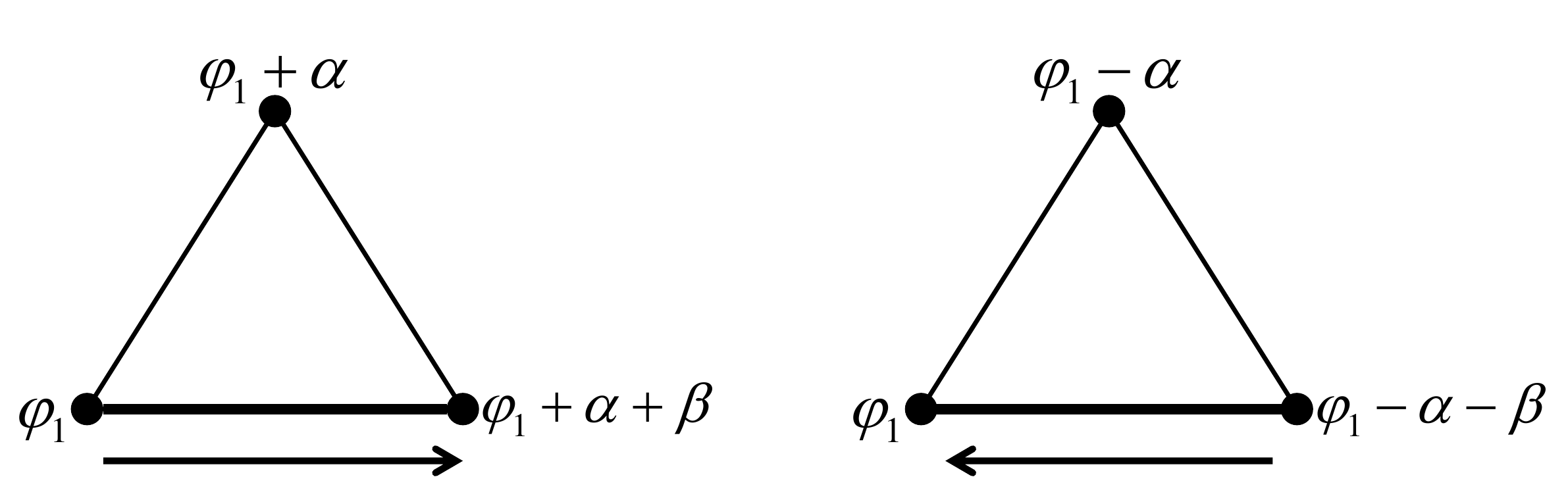}
\caption{Two angle configurations on triangle indicated by a right
and a left arrows.} \label{Fig_arrows}
\end{figure}

Comparing the latter equations with expressions
(\ref{function01}-\ref{function1}), we conclude that the condition
to have $6$ ground states (\ref{states0}-\ref{states}) and $2$
excited states (\ref{excited0},\ref{excited}) on the triangle
requires that $\varphi_2=\varphi_1\pm\alpha$ and
$\varphi_3=\varphi_2\pm \beta$. Thus, there are two possible angle
configurations on triangle:
\begin{eqnarray}
&&\{\varphi _{1},\varphi _{1}+\alpha ,\varphi _{1}+\alpha +\beta \}
\nonumber \\
&&\{\varphi _{1},\varphi _{1}-\alpha ,\varphi _{1}-\alpha -\beta \}
\end{eqnarray}%
with arbitrary $\varphi_1$. These two conditions can be marked by
arrows on AF bonds pointing in the direction of angle increase
(see Fig.\ref{Fig_arrows}). Then all allowed terms in wave
function (\ref{function}) can be represented graphically in a
similar way. Therefore, there is an one-to-one correspondence
between allowed terms in the function (\ref{function}) and allowed
configurations of the arrows. After this mapping the calculation
of the number of the ground states is equivalent to counting of
the allowed configurations of the arrows and this problem is
reduced to the solution of the corresponding lattice gas model. We
note that in this angle configuration approach (ACA) it is
impossible to separate the contributions to the ground state
degeneracy from the isolated localized magnons and from the bound
magnon complexes. However, the number of isolated magnons can be
found using the map to the hard-core representation
\cite{Derg,Zhitomirsky} and comparison of it with the total number
of allowed configurations gives the estimate of the contribution
of complexes.

As an example of an application of ACA approach, we calculate the
ground state degeneracy of the saw-tooth chain at the critical
point. There are two allowed configurations on each triangle and
$2^{n}$ configurations ($n=\frac{N}{2}$ is number of triangles) on
chain of triangles, and so, $2^{n}$ different wave functions
(\ref{function}). Each such function contains $N+1$ terms with
different $S_{tot}^{z}$. Hence, we have $W=(2n+1)2^{n}$ ground
state wave functions. It is interesting to compare this estimate
with the exact results for $W$ obtained in
\cite{Dmitriev,Dmitriev1}. Exact value $W$ is proportional to
$2^{n}$ though a pre-exponential factor $f(n)$ is different for
different special cases. For example, $f(n)=(\frac{n}{2}+1)$ for
$\alpha =\beta =\frac{\pi }{4}$ and $f(n)=(\frac{n}{3}+1)$ for
$\alpha =\beta =\frac{\pi }{3}$. For the isotropic model
$W=\sqrt{\frac{2n}{\pi }}2^{n}$. We see that the estimate of the
ground state degeneracy based on the wave function
(\ref{function}) and counting of the allowed arrow configurations
gives correct value for the exponent and thus, correctly
reproduces the value of the residual entropy.

We have to make two remarks concerning the wave function
(\ref{function}). The first one is related to the mixing in
Eq.(\ref{function}) of states with different $S^{z}$ subspaces.
Therefore, it is necessary to project wave function
(\ref{function}) to state with fixed $S^{z}$. The second remark
concerns the problem of linear independence of the states
corresponding to different arrow configurations in
(\ref{function}). These problems can be solved for finite systems,
as was performed for the special case $\alpha =\beta =\frac{\pi
}{3}$ in \cite{3-colors,3-colors1}, but it is not clear how to
solve them in the thermodynamic limit.

Nevertheless, the consideration of the F-AF saw-tooth chain shows
that the total number of degenerate ground states coincides with
exact $W$ up to preexponential factor. One can assume that the ACA
approach provides this accuracy for the two-dimensional models
considered in the next Section as well.

\section{Ground state degeneracy for the Tasaki and kagome lattices}

\begin{figure}[tbp]
\includegraphics[width=3in,angle=0]{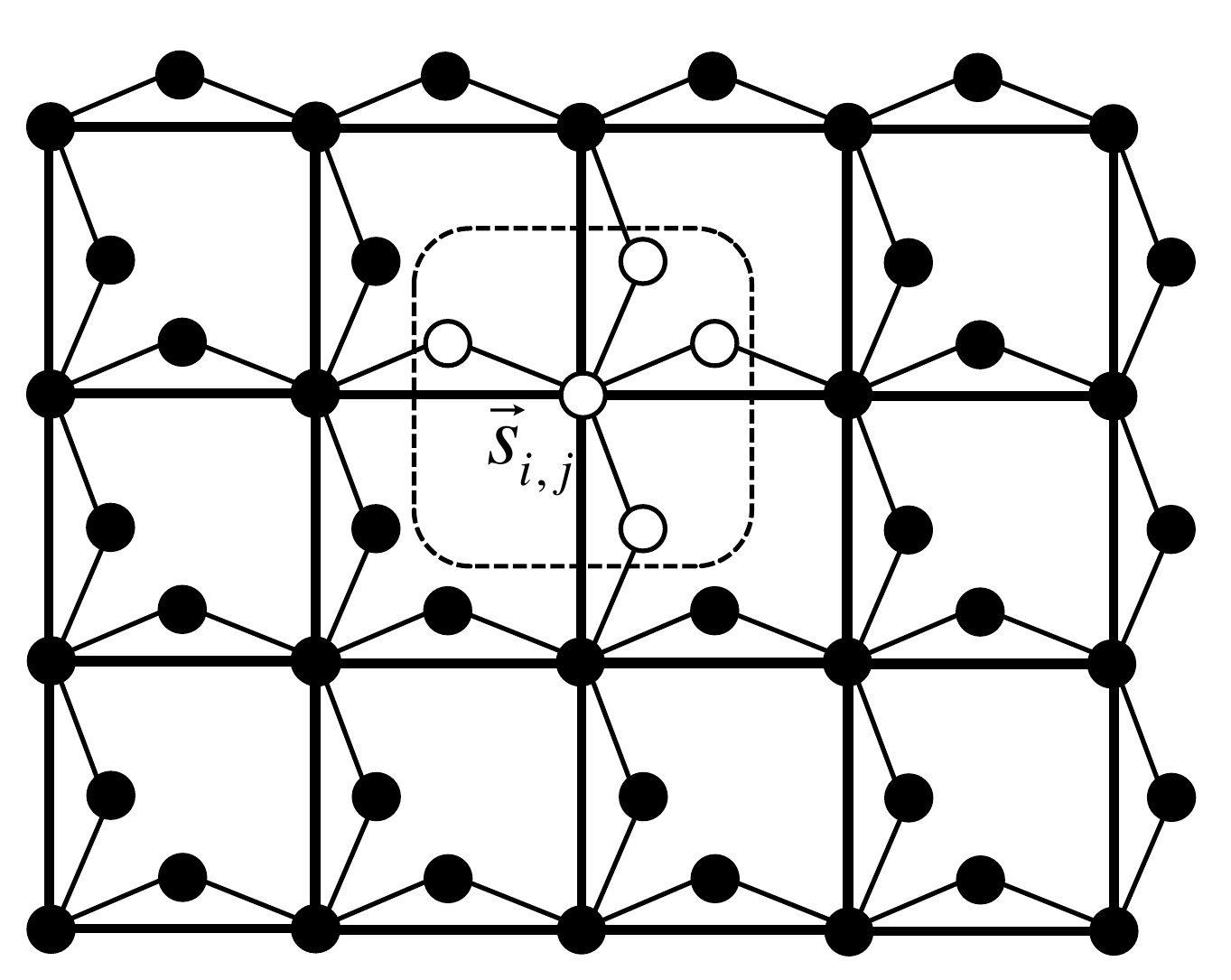}
\caption{Tasaki lattice with outlined trapping cell occupied by
the localized magnon on site ($i,j$).} \label{Fig_tasaki}
\end{figure}

In this Section we consider the F-AF models with the macroscopic
ground state degeneracy in the two-dimensional Tasaki lattice
\cite{Tasaki}. The Tasaki lattice is decorated square lattice
consisting of corner sharing triangles as it is shown in
Fig.\ref{Fig_tasaki}. The F-AF model on this lattice is a
generalization of the F-AF saw-tooth chain to the two-dimensional
case. The corresponding local Hamiltonian on each triangle has the
form (\ref{H1}), where interaction between the basal spins forming
square lattice is antiferrmagnetic and for simplicity we take
equal ferromagnetic bonds: $J_{1}=J_2$ and $\Delta _{1}=\Delta
_{2}$.

We consider this model at the transition points between the
ferromagnetic and other ground state phases (antiferromagnetic or
ferrimagnetic) when the F and AF interactions satisfy conditions
given in Section II. Then the model in zero-magnetic field has the
ground state consisting of the independent localized magnons and
the bound magnon complexes. The magnons are localized in trapping
cells and each trapping cell consists of one site of an
underlaying square lattice and four neighboring decorating sites,
as shown in Fig.\ref{Fig_tasaki}. The wave function of one
localized magnon on site ($i,j$) has a form
\begin{equation}
\hat{\varphi}_{i,j}\left\vert \downarrow \downarrow \ldots \downarrow
\right\rangle =[2\Delta _{1}s_{i,j}^{+}+s_{i,j\pm \frac{1}{2}}^{+}+s_{i\pm
\frac{1}{2},j}^{+}]\left\vert \downarrow \downarrow \ldots \downarrow
\right\rangle  \label{magnon}
\end{equation}

The bound magnon complexes contain overlapping localized magnons.
For example, two-magnon complex with the center on site ($i,j$) is
\begin{equation}
\hat{\varphi}_{i,j}[(2\Delta _{1}^{2}-1)\hat{\varphi}_{i,j}+\hat{\varphi}%
_{i,j\pm 1}+\hat{\varphi}_{i\pm 1,j}]\left\vert \downarrow \downarrow \ldots
\downarrow \right\rangle  \label{bound magnon}
\end{equation}

The number of the independent localized magnons can be found by
mapping to the hard-square model. The partition function of the
latter is known \cite{Baxter} and it is $Z=W_{0}\approx
1.5033^{N}$ ($N$ is number of squares in the lattice). However, it
is not the case for the contribution of magnon complexes to the
ground state manifold, because mapping them to any classical
lattice model is unknown. Therefore, we use the ground state
function (\ref{function}) and its arrow representation for an
estimate of $W$ .

\begin{table}[tbp]
\caption{Maximal eigenvalue $\lambda_L$ of transfer-matrix for
Tasaki stripe of width $L$ and the ratio $\lambda_L/\lambda_{L-1}$
in parenthesis.}
\label{tab:table1}%
\begin{ruledtabular}
\begin{tabular}{ccc}
\textrm{L} & \textrm{Case $(\alpha +\beta)\neq\frac{\pi}{2}$} &
{\textrm{Case $(\alpha
+\beta)=\frac{\pi}{2} $}}\\
\colrule
1 & 2 & 2 \\
2 & 3 (1.5) & 4 (2)\\
3 & 4.562 (1.521) & 8 (2)\\
4 & 6.972 (1.528) & 16 (2)\\
\end{tabular}
\end{ruledtabular}
\end{table}

\begin{figure}[tbp]
\includegraphics[width=4in,angle=0]{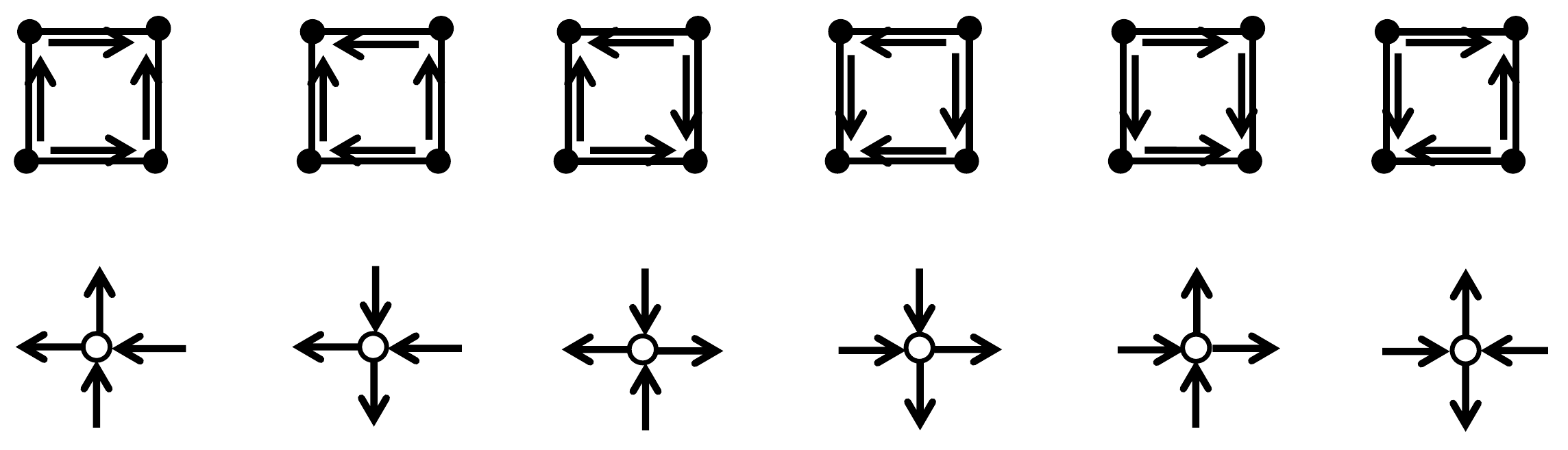}
\caption{Six allowed arrow configurations on elementary square
(top) of Tasaki lattice and six corresponding vertex
configurations (bottom).} \label{Fig_6conf}
\end{figure}

As shown in Fig.\ref{Fig_arrows}, the change of the angle along AF
bonds is $\pm (\alpha +\beta )$. Requirement that the angle change
along any closed loop on square lattice be zero leads to the
selection rule on the allowed configurations of arrows on each
elementary square: two arrows must be directed clockwise and two
arrows are directed counter-clockwise. This condition allows only
6 arrow configuration out of 16 total on each elementary square,
as shown in Fig.\ref{Fig_6conf}. The number of allowed
configurations on a Tasaki stripe of finite width $L$ and infinite
length $n$ ($n\gg 1$) is given by the largest eigenvalue, $\lambda
_{L}$, of the corresponding transfer-matrix, so that $W(L)=\lambda
_{L}^{n}$. The results for $\lambda _{L}$ for $L=1,2,3,4$ are
presented in Table 1 in the second column. For extrapolation of
these results to $L\to\infty $ we follow the ratio $\lambda
_{L}/\lambda _{L-1}$ and found that $\lambda _{L}/\lambda
_{L-1}\to\xi \approx 1.54$. This means that the ground state
degeneracy for large square lattice $L\times n$ behaves as $W=\xi
^{N}$ where $N=nL$.

It is worth noting that the above problem of counting of all
allowed arrow configurations on square lattice can be mapped to
the six-vertex model (square-ice model). For this purpose one
should consider each elementary square of Tasaki lattice as a new
vertex and rotate all arrows on $90^\circ$ clockwise (or
counterclockwise). As a result of this procedure $6$ allowed arrow
configurations on elementary square are mapped to the $6$ types of
vertex, as shown in Fig.\ref{Fig_6conf}. The partition function of
the six-vertex model was found by Lieb \cite{Lieb}:
$W=(4/3)^{3N/2}\approx 1.54^{N}$, which perfectly coincides with
our extrapolation value. As we can see from a comparison of $W$
and $W_{0}$ the magnon complexes give main contribution to the
ground state degeneracy in the thermodynamic limit, though for
finite systems the contribution of the independent localized
magnons can be essential. Taking into account that there are three
spins per site in the square lattice the residual entropy per spin
is $\mathcal{S}=\frac{1}{3}\ln \xi =0.144$.

Now we consider the special case $(\alpha +\beta )=\frac{\pi }{2}$
on Tasaki lattice. For this case the number of allowed
configurations is larger than for other cases. This is due to the
existence of two additional allowed arrow configurations in each
elementary square: all four arrows are directed clockwise or
counter-clockwise. These two extra configurations gives the change
of angles around the square equal to $\pm 2\pi $ for $(\alpha
+\beta )=\frac{\pi }{2}$. The largest eigenvalues of the
transfer-matrix for stripe of finite width $L$ for this case are
$\lambda _{L}=2^{L}$, as shown in Table 1, which means $\xi =2$.
This result is in agreement with that for the eight-vertex model,
where $W=2^{N}$ \cite{Baxter3}. The residual entropy per spin in
this case is $\mathcal{S}=\frac{1}{3}\ln 2=0.231$.

\begin{table}[tbp]
\caption{Ground state degeneracy of Tasaki stripe $2\times n$}
\label{tab:table2}%
\begin{ruledtabular}
\begin{tabular}{ccc}
\textrm{n} & \textrm{Isotropic model} & {\textrm{Case $(\alpha
+\beta)=\frac{\pi}{2} $}}\\
\colrule
2 & 52 & 64 \\
3 & 158 & 128 \\
4 & 632 & 1792 \\
5 & 1882 & 2048 \\
6 & 6884 & 40960 \\
\end{tabular}
\end{ruledtabular}
\end{table}

In order to verify the above ACA results we performed numerical
calculations of quantum spin model (\ref{H}). Unfortunately, the
size of the system accessible for numerical calculations on Tasaki
lattice is very limited, because Tasaki lattice contains three
spins per site. Therefore, we had to restrict ourselves to exact
diagonalization (ED) of Tasaki stripes $2\times n $ with $n\leq
6$. The results are presented in Table 2. As follows from Table 2,
the ground state degeneracy for the special case $(\alpha +\beta
)=\frac{\pi }{2}$ is higher than that for the isotropic case. The
isotropic case can be considered as the general case, because all
other cases of $\alpha $ and $\beta $ (except the case $(\alpha
+\beta )=\frac{\pi }{2}$) have the same or very close ground state
degeneracy. The extrapolation of the data for the isotropic case
shows the exponent $\lambda _{2}\simeq 3\div 3.5$, which is in a
good agreement with the estimate $\lambda _{2}=3$, presented in
Table 1 for the general case. This agreement allows us to expect
that the ACA result $W=(4/3)^{3N/2}$ for the general case gives a
good approximation of the ground state degeneracy on Tasaki
lattice.

The higher ground state degeneracy for the special case $(\alpha
+\beta )=\frac{\pi }{2}$ can be explained by the following
property of the model with the Hamiltonian $H_{S,1}$
(\ref{sqrt2}). As follows from Eq.(\ref{bound magnon}) the magnon
states $\varphi _{i,j}$ for this model can be located on any sites
of the lattice including nearest neighbors. That is the two magnon
states $\varphi _{i,j}\varphi _{i+1,j}$ and $\varphi _{i,j}\varphi
_{i,j+1}$ are exact ground states at $(\alpha +\beta )=\frac{\pi
}{2}$. This allows to find the ground state degeneracy exactly.
For two-dimensional Tasaki model comprised of $N$ squares the
exact number of ground states $G(N,k)$ in the spin sector
$S_{tot}^{z}=\frac{3N}{2}-k$ reduces to the number of possible
combinations to place $k$ local magnon functions $\varphi _{i,j}$
in $N$ sites. It is given by the binomial coefficients
$G(N,k)=C_{N}^{k}$ and, therefore, the total degeneracy is
$W=2\cdot 2^{N}$ (the factor $2$ comes from two sectors
$S^z_{tot}>0$ and $S^z_{tot}<0$ giving equal contribution). This
is a correct result for any size of Tasaki lattice $L\times n$
except the case when both $n$ and $L$ are even numbers. In this
case the exact number of ground states $G(N,k)$ is
\begin{equation}
G(N,k)=\sum_{m=0,2\ldots }^{k}C_{N}^{m}  \label{G1}
\end{equation}%
for even $k$ and%
\begin{equation}
G(N,k)=\sum_{m=1,3\ldots }^{k}C_{N}^{m}  \label{G2}
\end{equation}%
for odd $k$. So that the total degeneracy is
\begin{equation}
W=\sum_{k}G(N,k)=2^{N}\left( N+1\right)   \label{W}
\end{equation}

For Tasaki stripe $2\times n$ the total degeneracy has slightly
different expression: $W=2^{2n-1}(3n+2)$. It is caused by the
different number of spins per square: $3$ apical and $2$ basal
spins for Tasaki stripe and $2$ apical and $1$ basal spin for
Tasaki lattice.

In the thermodynamic limit all results for $W$ gives the same
residual entropy per spin $\mathcal{S}=\frac{1}{3}\ln 2$. The
above expressions are confirmed by ED results for $2\times n$
stripe presented in Table 2 and on Tasaki lattices $3\times 3$,
$3\times 4$, $4\times 4$.

Another exact result for the model (\ref{sqrt2}) is related to the
spontaneous magnetization at $T=0$. The partition function $Z$
from the degenerate ground state in the magnetic field $h$ is
\begin{equation}
Z=2\sum_{k}G(N,k)\cosh [(\frac{3}{2}N-k)\frac{h}{T}]  \label{Z}
\end{equation}

The magnetization per spin $m$ is given by conventional
thermodynamic equation:
\begin{equation}
m=\frac{T}{3N}\frac{\partial\ln Z}{\partial h}  \label{magn1}
\end{equation}

The calculation of $Z$ results in the expression for the
magnetization at $N\gg 1$ in a form
\begin{equation}
m=\frac{3+\exp (-h/T)}{6(1+\exp (-h/T))}  \label{magn}
\end{equation}

At $h/T\to 0$ the spontaneous magnetization is $m=\frac{1}{3}$. It
means that the ground state is magnetically ordered and the
magnetization is changed from $m=\frac{1}{2}$ in the ferromagnetic
phase to $m=\frac{1}{3}$ on the phase boundary. Unfortunately, we
can not calculate the magnetization for the cases $(\alpha +\beta
)\neq \frac{\pi }{2}$ but we believe that the spontaneous
magnetization exists for the general case of model (\ref{H}).

\begin{figure}[tbp]
\includegraphics[width=3in,angle=0]{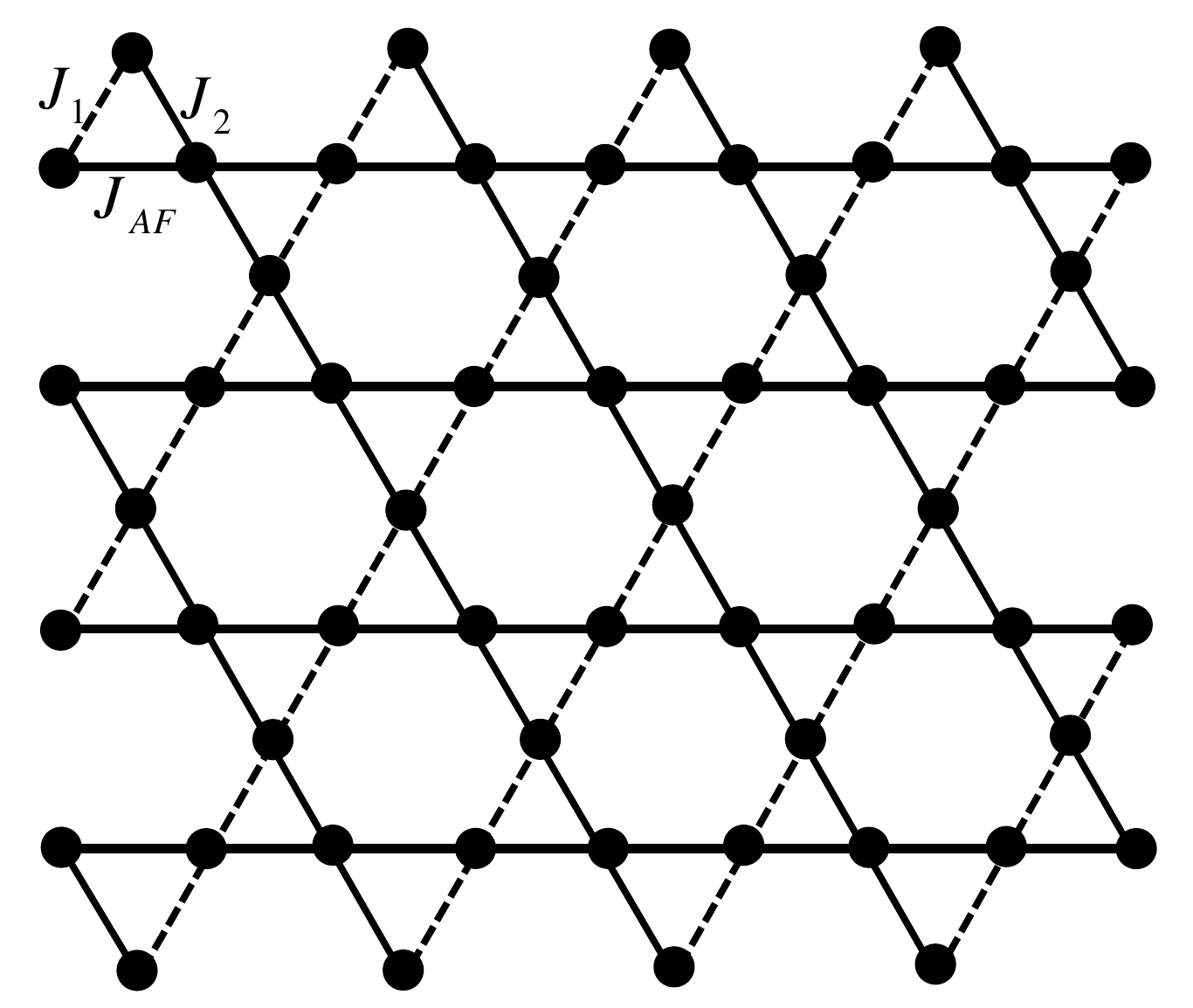}
\caption{Kagome lattice with three different type of interactions.}
\label{Fig_kagome}
\end{figure}

\begin{figure}[tbp]
\includegraphics[width=3in,angle=0]{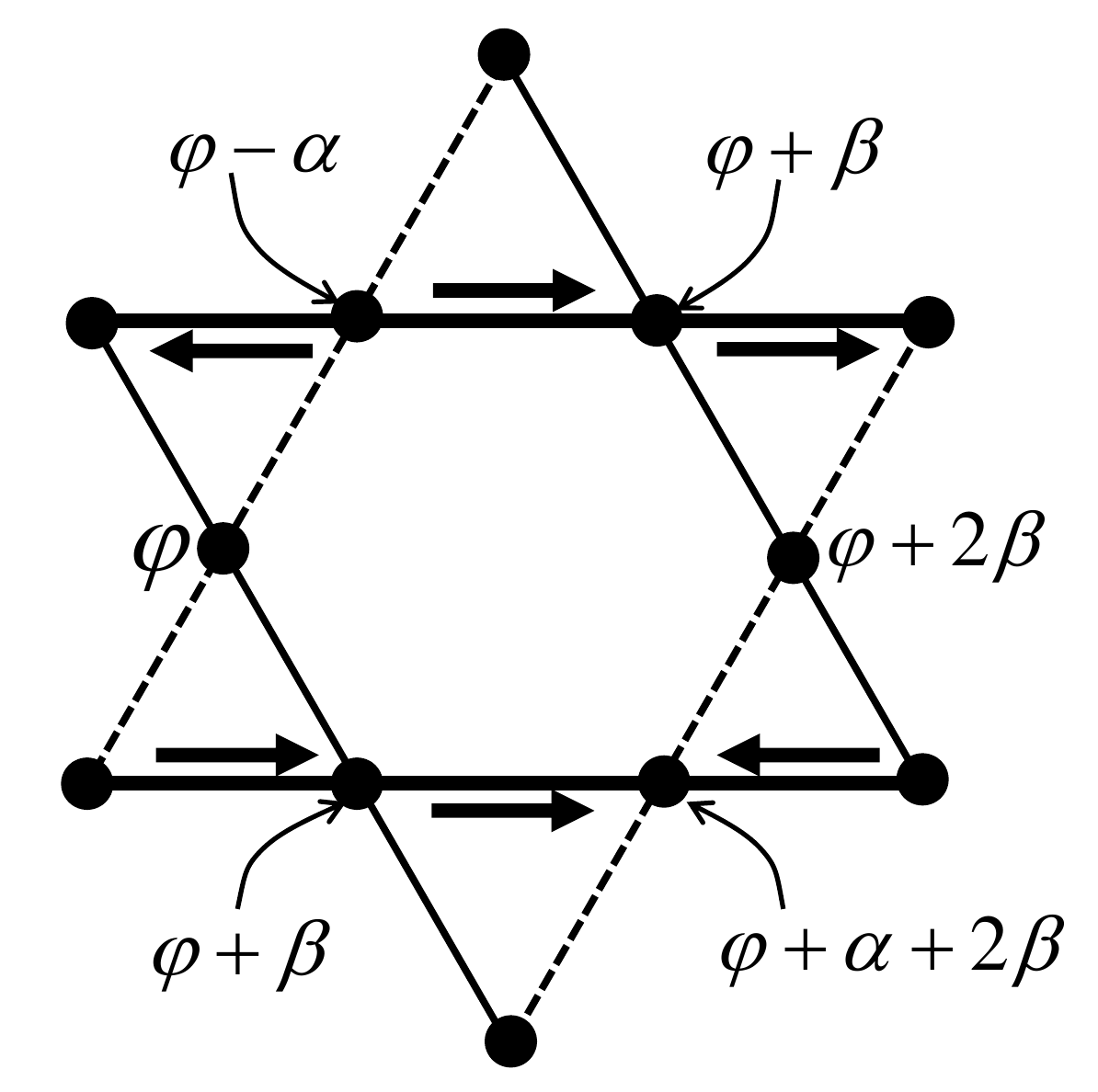}
\caption{Example of possible arrow configuration around one
hexagon and corresponding angles on hexagon sites.}
\label{Fig_david}
\end{figure}

Thus, the ACA approach correctly distinguishes the general case
and the special case $(\alpha +\beta )=\frac{\pi }{2}$ on Tasaki
lattice, gives a good approximation for the ground state
degeneracy in the general case and reproduces the exact result for
the special case. Unfortunately, the ACA approach does not work in
general case for the F-AF model (\ref{H}) on the kagome lattice
shown in Fig.\ref{Fig_kagome}. The point is that the total number
of arrow configuration is strongly reduced by the self-consistency
condition of zero angle change around each hexagon of kagome
lattice, similar to the selection rule for elementary square on
Tasaki lattice. The angle distribution of each hexagon is governed
by the arrow configuration on six adjacent triangles. The total
number of arrow configurations on six adjacent triangles is
$2^6=64$. However, for general choice of $\alpha$ and $\beta$ only
$10$ arrow configurations are allowed, one of which is shown in
Fig.\ref{Fig_david}.

\begin{figure}[tbp]
\includegraphics[width=3in,angle=0]{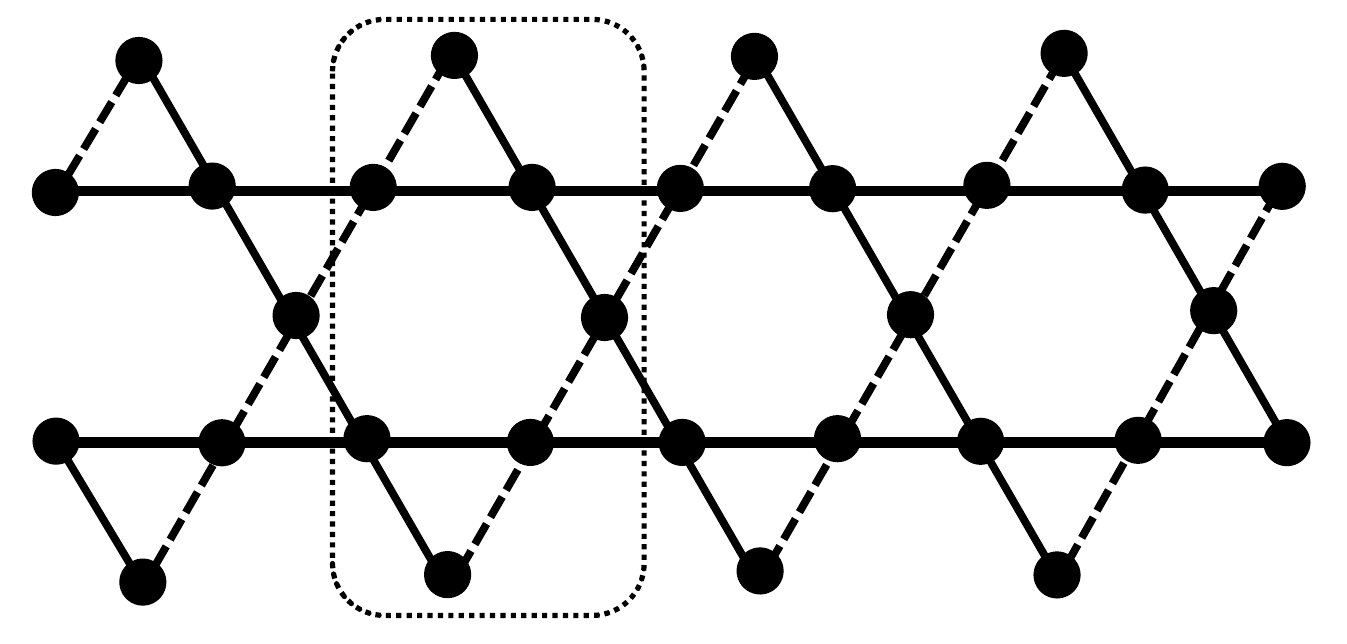}
\caption{Kagome stripe with outlined elementary cell.}
\label{Fig_kagome_stripe}
\end{figure}

The calculation of the number of the allowed arrow configurations
on kagome lattice is very complicated problem. Therefore, at first
we consider the kagome stripe of width $L=2$ shown in
Fig.\ref{Fig_kagome_stripe}. The maximal eigenvalue of the
corresponding transfer-matrix is $\lambda_2=3$, so that the number
of allowed arrow configurations on the kagome stripe is $W=3^{n}$,
where $n$ is the number of elementary cells containing $7$ spins
(see Fig.\ref{Fig_kagome_stripe}). Similar calculations for the
kagome stripe of width $L=3$ leads to $\lambda_3=2.618$ and we
also note that the kagome stripe of width $L=1$ is nothing but the
saw-tooth chain with $\lambda_1=4$ (three spins per elementary
cell). Therefore, we see that $\lambda_L$ decreases with the
increase of $L$ and tends to some value $\lambda_{\infty}$, which
implies that for infinite kagome lattice ($L\to\infty $) the
number of arrow configurations behaves as
$W=\lambda_{\infty}^{n}$. Such dependence leads to zero residual
entropy per spin $\mathcal{S}=\ln\lambda_{\infty}/L \to 0$.

There are several special cases ($\alpha =\beta$,
$\alpha+\beta=\frac{\pi}{2} $, $\alpha+2\beta=\pi$,$\ldots$),
which have higher number of allowed arrow configurations per
hexagon caused by commensurability of $\alpha$ and $\beta$ and/or
the possibility for the angle change around the hexagon to be $\pm
2\pi $. However, the diagonalization of the corresponding
transfer-matrix for kagome stripe of widths $L=1,2,3$ showed that
the maximal eigenvalues $\lambda_L$ do not increase with the
increase of $L$ and, therefore, there is no residual entropy. The
only exception is the special case $\alpha =\beta =\frac{\pi}{3}$
with local Hamiltonian given by Eq.(\ref{pi/3}), which can be
described by the three coloring approach
\cite{3-colors,3-colors1}. In this case the number of allowed
arrow configurations per hexagon is $22$ and the calculated
eigenvalues are $\lambda_1=4$, $\lambda_2=5.562$, $\lambda_3=7.894
$. The ratios $\lambda_2/\lambda_1=1.39$ and
$\lambda_3/\lambda_2=1.419$ converge to the exact result
$\xi=1.461$ \cite{Baxter1}, so that the total number of arrow
configurations is $W\approx 1.461^{N}$ for the kagome-lattice
containing $N$ hexagons. This value should be compared with the
number of the isolated magnons $W_{0}$, which is given by the hard
hexagon model having $W_{0}\approx 1.395^{N}$ configurations
\cite{Baxter2}. Therefore, we conclude that the main contribution
to the ground state degeneracy in this case is given by the magnon
complexes again.

\begin{figure}[tbp]
\includegraphics[width=4in,angle=0]{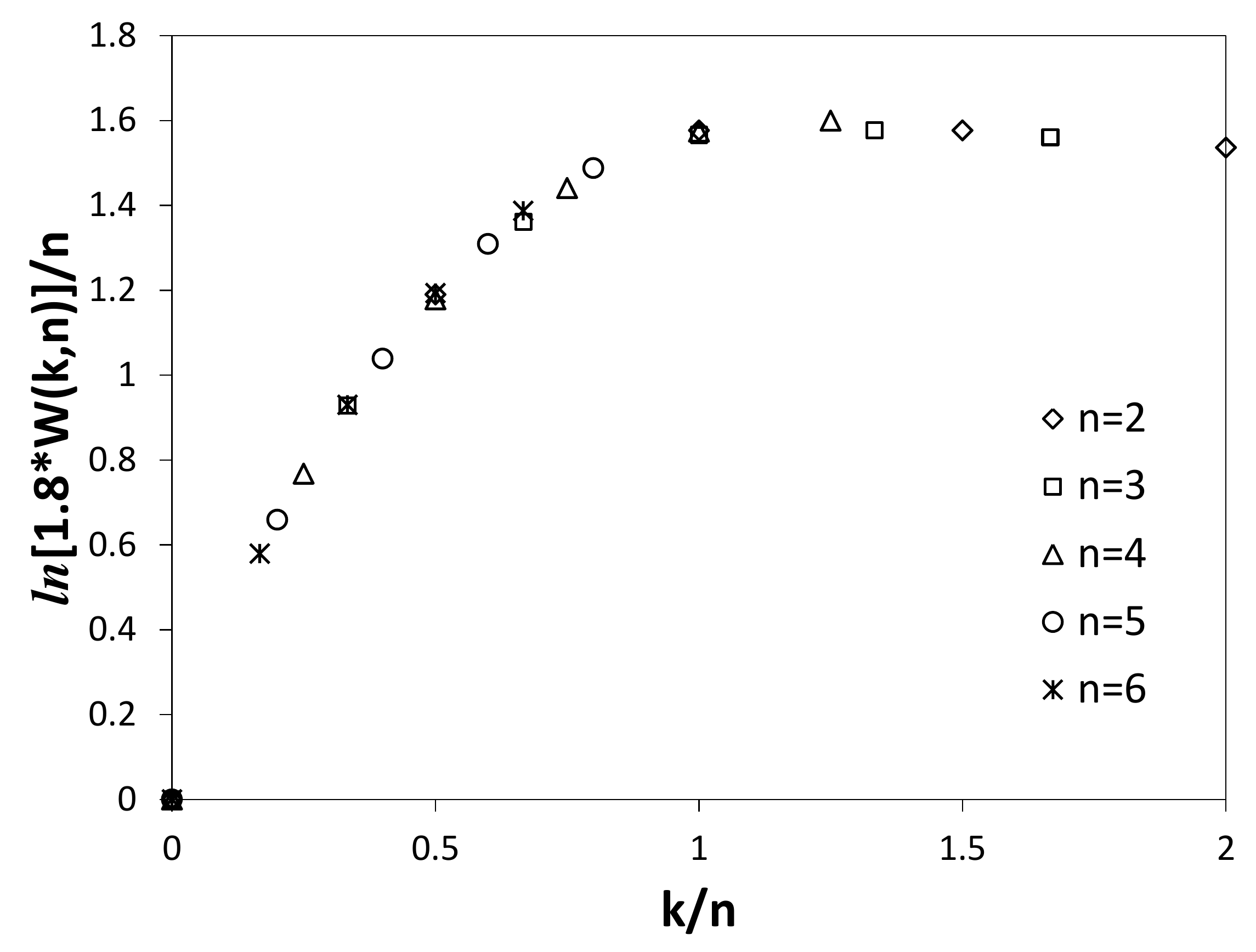}
\caption{Dependence of number of ground states $W(k,n)$ on number
of magnons $k$ for kagome stripe of width $L=2$ and length $n$ for
the case $\alpha =\beta =\frac{\pi}{3}$.}
\label{Fig_W_k_kagome_stripe}
\end{figure}

We performed ED calculations for some special cases (including the
isotropic version) of the quantum spin F-AF model (\ref{H}) on the
kagome-stripe of width $L=2$ and length $n\leq 6$
(Fig.\ref{Fig_kagome_stripe}). The obtained data for the number of
ground states $W(k,n)$ in different sectors of $S^z=S^z_{\max}-k$
for the case $\alpha =\beta =\frac{\pi}{3}$ are plotted in
Fig.\ref{Fig_W_k_kagome_stripe} as $\ln(1.8W(k,n))/n$ vs. $k/n$.
(Here the factor $1.8$ is fitting parameter, which describes $1/n$
corrections.) We see that the data for different $k$ and $n$ lies
perfectly on one curve, which justifies the law $W(k,n)\sim
\exp[nf(k/n)]$. The total number of ground states are defined by
the maximum of this dependence, which takes place at $k\sim 1.2n$
and has the value $\simeq 1.6$. This allows to determine the
thermodynamic behavior $W \sim \exp(1.6n)\simeq 5^n$, which is in
a good agreement with the estimate of ACA approach for this case
$\lambda_2=5.562$.

The same treatment of the numerical data for the case $\alpha
=\beta =\frac{\pi }{4}$ and for the isotropic model leads to the
results $W\simeq 5^{n}$ and $W\simeq 4.5^{n}$, respectively.
Unfortunately, the ED calculations of the F-AF model for the
kagome stripes with $L>2$ and rather large $n$ are not accessible.
However, the fact that the number of ground states for kagome
stripe with $L=2$ grows faster than $4^{n}$, which corresponds to
$L=1 $ (saw-tooth chain), is a strong argue in favor of the fact
that the considered F-AF models on the kagome lattice have the
macroscopic ground state degeneracy. Moreover, we can give the
lower bound for $W$ demonstrating that $\xi >1$. Because $W>W_{0}$
the lower bound for $W$ is the number of the isolated magnons
$W_{0}$. The value of $W_{0}$ can be found using the map of the
considered models on the kagome lattice onto the lattice gas of
hard-core particles. For the case $\alpha +\beta \neq \frac{\pi
}{2}$ it is a hard-hexagon model (the triangular lattice with
nearest-neighbor exclusion) and $W_{0}=1.395^{N}$ \cite{Baxter2}.
For the case $\alpha =\beta =\frac{\pi }{4}$ it can be shown that
the nearest-neighbor exclusion does not act between
nearest-neighbor sites in the same stripe and the corresponding
lattice-gas model reduces to the hard-square lattice. In this case
$W_{0}=1.503^{N}$ \cite{Baxter2}. Therefore, for all considered
models $\xi >1$.

\section{Summary}

In this paper we study the ground state degeneracy of frustrated
quantum spin-$\frac{1}{2}$ F-AF model at the critical ratio of the
ferromagnetic and antiferromagnetic interactions corresponding to
the boundaries of the ground state phase diagram. We constructed
the general form of the anisotropic Hamiltonian acting on a
lattice consisting of corner sharing triangles, the ground state
of which is macroscopically degenerated in zero magnetic field.
This general form of Hamiltonian is parameterized by two angles
$\alpha$ and $\beta$.

To calculate the number of the degenerate ground states we use the
wave function of the specific form (\ref{function}). This function
can be represented in terms of allowed arrow configurations on the
corresponding lattice and the number of these configurations gives
the estimate of the ground state degeneracy. This ACA approach
reproduces the known exact result for the F-AF saw-tooth chain in
the critical points with the exponential accuracy.

For two-dimensional Tasaki model the ACA approach correctly
distinguishes the general case and the special case $(\alpha
+\beta )=\frac{\pi }{2}$ giving different estimate of the ground
state degeneracy, $W=1.5396^N$ for the general case and $W=2^N$
for the special case $(\alpha +\beta )=\frac{\pi }{2}$. Performed
exact diagonalization calculations on finite parts of Tasaki
lattice confirm macroscopic degeneracy of the ground state and
show a good agreement with ACA results for the general case. For
the special case $(\alpha +\beta )=\frac{\pi }{2}$ the ground
state degeneracy is calculated exactly, it is $W\sim 2^N$ at
$N\to\infty$ as correctly predicted by ACA approach.

The exact diagonalization of finite kagome lattices indicate the
macroscopic ground state degeneracy for the general case on the
kagome lattice. However, the ACA approach does not predict it.
This is due to the fact that the number of allowed arrow
configurations is strongly limited by the condition of zero total
angle change around each hexagon of the kagome lattice. The only
case where the ACA predicts macroscopic ground state degeneracy is
the special case $\alpha=\beta=\frac{\pi}{3}$, where the above
constraint is not so severe and the ACA reproduces the exact
result $W=1.461^N$. Comparing the kagome and Tasaki lattices, we
assume that the ACA approach can correctly predict ground state
degeneracy for such lattices where each triangle has one spin that
does not simultaneously belong to an adjacent triangle.

The ground state manifold for all above cases and lattices
consists of isolated magnons and the bound magnon complexes,
$W=W_0+W_b$. The number of the independent localized magnons can
be found by the mapping to the hard-square model with
$W_{0}\approx 1.503^{N}$ for Tasaki lattice and to the
hard-hexagon model with $W_{0}\approx 1.395^{N}$ for kagome
lattice. As follows from the comparison of $W$ and $W_{0}$, the
magnon complexes give always main contribution to the ground state
degeneracy in the thermodynamic limit, though for finite systems
the contribution of the independent localized magnons can be
essential.

The special case $\alpha=\beta=\frac{\pi}{4}$, for which the
Hamiltonian takes a simple form (\ref{sqrt2}), requires a separate
comment. In this case the local magnon states can be located on
any sites of the Tasaki lattice including nearest neighbors. This
leads to higher ground state degeneracy and allows to calculate it
exactly. It is obvious that this special model can be extended to
three-dimensional Tasaki lattice with similar exact result for the
ground state degeneracy.

In this paper we paid attention only to the ground state
degeneracy. However, as was shown for the saw-tooth chain
\cite{Dmitriev1}, the spectrum of the studied models can have
interesting and unusual features: exponentially low excitations
and multi-scale hierarchy. Exact diagonalization of finite Tasaki
and kagome systems indicates the presence of such extremely low
excitations. However, the accessible for ED systems are too small
for the quantitative analysis of the spectrum. This interesting
problem requires further study.


\begin{thebibliography}{99}
\bibitem{Lacroice} C.~Lacroix, P.~Mendels and F.~Mila, eds., Introduction to
frustrated magnetism. Materials, Experiments, Theory
(springer-Verlag, Berlin, 2011).

\bibitem{Diep} H.~T.~Diep (ed) 2013 Frustrated Spin Systems (Singapore; World
Scientific).

\bibitem{ModernPhysics} O.~Derzhko, J.~Richter, M.~Maksymenko, Int.~J.~Mod.~Phys B
\textbf{29}, 153007 (2015).

\bibitem{Derg} O.~Derzhko, J.~Richter, Phys.~Rev. B \textbf{70}, 104415 (2004).

\bibitem{Zhitomirsky} M.~E.~Zhitomirsky, H.~Tsunetsugu, Phys.~Rev. B \textbf{70}, 100403(R)
(2004).

\bibitem{Tsun} M.~E.~Zhitomirsky, H.~Tsunetsugu, Progr.~Theor.~Phys.~Suppl.
\textbf{160}, 361 (2005).

\bibitem{Derg1} O.~Derzhko, J.~Richter, Eur.~Phys.~J. B \textbf{52}, 23 (2006).

\bibitem{Zhit} M.~E.~Zhitomirsky, H.~Tsunetsugu, Phys.~Rev. B \textbf{75},
224416 (2007).

\bibitem{Schulen} J.~Schnack, J.~Schulenberg, J.~Richter, Phys.~Rev. B \textbf{98}, 094423 (2018).

\bibitem{Dmitriev} V.~Ya.~Krivnov, D.~V.~Dmitriev, S.~Nishimoto, S.~-L.~Drexsler,
J.~Richter, Phys.~Rev. B \textbf{90}, 014441 (2014).

\bibitem{Gd} A.~Baniodeh, N.~Magnani, Y.~Lan, G.~Buth, C.~E.~Anson, J.~Richter,
M.~Affronte, J.~Schnack, A.~K.~Powell, Quantum Matter, \textbf{3},
10 (2018).

\bibitem{DKRS} D.~V.~Dmitriev, V.~Ya.~Krivnov, J.~Richter, J.~Schnack, Phys.~Rev. B \textbf{99},
094410 (2019); Phys.~Rev. B \textbf{101}, 054427 (2020).

\bibitem{cooling} M.~E.~Zhitomirsky and A.~Honecker, J.\ Stat.\ Mech.: Theory and Experiment
\textbf{2004}, P07012 (2004).

\bibitem{kagomestripe} D.~V.~Dmitriev and V.~Ya.~Krivnov, J.\ Phys.: Condens.\
Matter \textbf{29}, 215801 (2017).

\bibitem{3-colors} H.~J.~Changlani, D.~Kochkov, K.~Kumar, B.~K.~Clark, E.~Fradkin,
Phys.~Rev.~Lett. \textbf{120}, 117202 (2018).

\bibitem{3-colors1} H.~J.~Changlani, S.~Pujari, C.~-M.~Chungr, B.~K.~Clark,
Phys.~Rev. B \textbf{99}, 104433 (2019).

\bibitem{Dmitriev1} D.~V.~Dmitriev, V.~Ya.~Krivnov, Phys.~Rev. B \textbf{92},
184422 (2015).

\bibitem{Johann} O.~Derzhko, J.~Schnack, D.~V.~Dmitriev, V.~Ya.~Krivnov,
J.~Richter, Eur.~Phys.~J. B \textbf{93}, 161 (2020).

\bibitem{ferri} D.~V.~Dmitriev and V.~Ya.~Krivnov, J.\ Phys.: Condens.\
Matter \textbf{28}, 506002 (2016); V.~Ya.~Krivnov and
D.~V.~Dmitriev, Russian J.\ Phys.\ Chem. B, \textbf{15}, 89
(2021).

\bibitem{kosoi} D.~V.~Dmitriev and V.~Ya.~Krivnov, J.\ Phys.: Condens.\
Matter \textbf{30}, 385803 (2018).

\bibitem{Tasaki} H.~Tasaki, Phys.~Rev.~Lett. \textbf{69}, 1608 (1992).

\bibitem{Baxter} R.~J.~Baxter, Ann. Combinatorics \textbf{3}, 191 (1999).

\bibitem{Lieb} E.~H.~Lieb, Phys.~Rev., \textbf{162}, 162 (1967).

\bibitem{Baxter1} R.~J.~Baxter, J.~Math.~Phys. (N.J.) \textbf{11}, 784 (1970).

\bibitem{Baxter2} R.~J.~Baxter, S.~K.~Tsang, J.~Phys A \textbf{13}, 1023 (1980).

\bibitem{Baxter3} R.~J.~Baxter, Exactly Solved Models in Statistical Mechanics
(Academic, London, 1982).

\end{thebibliography}
\end{document}